\documentclass[12pt,preprint]{aastex}

\newcommand{\xmm}{{\it XMM-Newton}}
\newcommand{\rosat}{{\it ROSAT}}
\newcommand{\einstein}{{\it Einstein}}
\newcommand{\hi}{H~{\small I}}

\newcommand{\ha}{H$\alpha$}
\newcommand{\sii}{[\ion{S}{2}]}
\newcommand{\snr}{SNR~0450$-$70.9}
\begin{document}

\title{Supernova Remnants in the Magellanic Clouds IV: 
X-ray Emission from the Largest SNR in the LMC}
\author{R. M. Williams\altaffilmark{1}, Y.-H. Chu\altaffilmark{1}, 
	J. R. Dickel, R. A. Gruendl\altaffilmark{1}}
\affil{University of Illinois at Urbana-Champaign, 1002 W. Green St., 
	Urbana, IL 61801}
\email{rosanina@astro.uiuc.edu, chu@astro.uiuc.edu, johnd@astro.uiuc.edu,
	gruendl@astro.uiuc.edu}
\author{R. Shelton}
\affil{Department of Physics and Astronomy, University of Georgia,
Athens, GA 30602-2451}
\email{rls@hal.physast.uga.edu}
\and
\author{S. D. Points, R. C. Smith}
\affil{Cerro-Tololo Inter-American Observatory, Casilla 603, La Serena, Chile}
\email{spoints@ctio.noao.edu, csmith@ctio.noao.edu}
\altaffiltext{1}{
Visiting Astronomer, Cerro Tololo Inter-American Observatory,
National Optical Astronomy Observatories, operated by the
Association of Universities for Research in Astronomy, Inc. (AURA)
under a cooperative agreement with the National Science Foundation.}

\begin{abstract}
We present the first X-ray detection of \snr, the largest known supernova 
remnant (SNR) in the Large Magellanic Cloud.  To study the physical 
conditions of this SNR, we have obtained \xmm\ X-ray observations, optical 
images and high-dispersion spectra, and radio continuum maps. Optical 
images of SNR 0450$-$70.9 show a large, irregular elliptical shell with 
bright filaments along the eastern and western rims and within the 
shell interior.  The interior filaments have higher [S II]/H$\alpha$ 
ratios and form an apparent inner shell morphology.  The X-ray emission 
region is smaller than the full extent of the optical shell, with the
brightest X-ray emission found within the small interior shell and on the 
western rim of the large shell.  The expansion velocity of the small shell 
is $\sim$220 km~s$^{-1}$, while the large shell is $\sim$120 km~s$^{-1}$.  
 
The radio image shows central brightening and a fairly flat radio spectral 
index over the SNR. However, no point X-ray or radio source corresponding 
to a pulsar is detected and the X-ray emission is predominantly thermal.  
Therefore, these phenomena can be most reasonably explained in terms of the 
advanced age of the large SNR.  Using hydrodynamic models combined with a 
nonequilibrium ionization model for thermal X-ray emission, we derived a 
lower limit on the SNR age of about 45,000 yr, well into the later stages
of SNR evolution.  Despite this, the temperature and density derived from 
spectral fits to the X-ray emission indicate that the remnant is still 
overpressured, and thus that the development is largely driven by hot gas 
in the SNR interior.

\end{abstract}


\section{Introduction}

Supernova remnants (SNRs) are the main contributors to the hot 
ionized component of the interstellar medium; however, the physical 
conditions and amounts of the hot gas in SNRs at late evolutionary 
stages are not well known. Theoretical models of hot gas in SNRs of 
ages 10$^5$--10$^6$ yr \citep[e.g.,][]{SC92,SC93,S99} cannot be 
compared easily to observations because few evolved SNRs are known 
and they are expected to be faint and difficult to observe.

\snr\ \citep{M+85} is the largest known remnant in
the Large Magellanic Cloud (LMC).  Its optical extent of
6\farcm5$\times$4\farcm7 corresponds to 98$\times$70 pc at
the LMC distance of 50 kpc \citep{F99}.  This SNR is not
near any known OB association or bright \ion{H}{2} region, and
the surface density of field OB stars is low in its vicinity
\citep{CK88}; thus, it is unlikely to be a case of a SNR
brightening a superbubble through an internal collision with
the walls of the shell of the superbubble \citep{CM90}.

While larger SNR candidates in the LMC have been identified at X-ray 
wavelengths \citep[e.g.,][]{C+00,Lowryetal04}, such candidates are 
in a tenuous medium, so that no optical or radio counterparts can be 
detected to confirm their physical nature.  \snr, on the other hand, 
has a \sii/\ha\ ratio of 0.7 and a radio spectral index estimated 
at $-$0.2 ($S_\nu \propto \nu^\alpha$), which clearly confirm its 
identity as an SNR \citep{M+85,CLM76}. The X-ray emission from \snr\ has 
never been detected; no pointed \einstein\ or \rosat\ observations
of this remnant were ever made.  While the \rosat\ All-Sky Survey did
cover this region, no emission from the SNR was detected.

To study the physical conditions of this large and evolved
SNR, we have obtained and analyzed \xmm\ X-ray observations, 
optical images and high-dispersion spectra, and additional 
radio continuum maps.  In this paper, \S2 describes these
observations, \S3 presents the results of our analysis, and
\S4 discusses the implications of our findings.

\section{Observations}

X-ray observations of the hot gas in \snr\ were made by \xmm\ in 
2001 November (Observation IDs 89210601 and 89210801). Unfortunately, 
the European Photon Imaging Camera (EPIC) PN camera was not turned on 
in the first observation, and only the Reflection Grating Spectrometer 
was used in the second.  Only the EPIC MOS data from the first 
observation were useful; the exposure time was 62.5 ks.

We received the pipeline-processed data from the \xmm\ Science Operations 
Centre (SOC).  Initial reduction and analyses were carried out using the 
Science Analysis Software (SAS) package provided by the SOC.  We filtered 
out data with high background or poor event grades, leaving 48.8 ks of good
exposure time for each of the two EPIC MOS detectors.  Images and spectra 
were then extracted from the filtered event files.  The EPIC MOS images 
were combined, using the experimental SAS task ``merge", to increase the 
signal-to-noise ratio.  The intrinsic on-axis point-spread function (PSF) 
of the telescopes associated with the MOS detectors are 4\farcs3 and 
4\farcs4 at 1.5 keV; our images are adaptively smoothed, further reducing 
the resolution in order to bin photons to a signal-to-noise ratio of 6 
(SAS asmooth task).  Source-free regions surrounding the SNR were used 
to produce background spectra, which were scaled and subtracted from the 
SNR spectra.  Further spatial and spectral analysis was done using the 
FTOOLS and XSPEC software. Spectra were binned to a minimum of 20 counts 
per bin to improve statistics. The spectra of the two EPIC MOS detectors 
were fitted jointly to produce the spectral parameters.

To map the dense swept-up shell of \snr, we used the \ha\ 
($\lambda$6561) + [\ion{N}{2}] ($\lambda\lambda$6548,6583),
\sii\ ($\lambda\lambda$6716,6731), and red continuum 
($\lambda_0$ = 6850\AA, $\Delta\lambda$ = 95\AA) images taken with 
the Curtis Schmidt Telescope at Cerro Tololo Inter-American Observatory 
(CTIO) as part of the Magellanic Cloud Emission-Line Survey
\citep[MCELS,][]{Smithetal99}. The exposure times were 300s, 600s
and 300s, respectively.  The images have $\sim3$\arcsec\ resolution.
All images were flux-calibrated.  Contribution from the [\ion{N}{2}]
lines is present in the \ha\ filter image, but due to the low
[\ion{N}{2}]/\ha\ ratios observed for older SNRs in the LMC 
\citep[0.2--0.3;][]{DS76} this contribution is expected to be small.
The red continuum image was scaled and 
subtracted from the \ha\ and \sii\ images to remove the stellar emission.  
The \sii\ and \ha\ images were clipped at $3\sigma$ and divided 
to make an \sii/\ha\ ratio map.

To examine the dynamic properties of this SNR, high-dispersion 
long-slit spectroscopic observations of the \ha\ 
and [\ion{N}{2}] $\lambda\lambda$6548, 6583 lines were obtained with 
the echelle spectrograph on the 4m Blanco Telescope at CTIO on 2000
December 6.  The detailed observing configuration can be found in the 
paper by \citet{Chuetal03}.  Briefly, the data array samples $\sim3$\arcmin\ 
along the slit with a pixel size of 0\farcs26, and covers $\sim80$ \AA\ 
along the dispersion axis with a pixel size of 0.082 \AA. A 1\farcs64 
slit width was used and the resulting instrumental profile has a full
width at half maximum (FWHM) of $13.5\pm0.5$ km~s$^{-1}$ at the \ha\ line.
The echelle observation was made with an E-W oriented slit for 
an exposure time of 1200 s.

Finally, to study the synchrotron radiation from \snr\ we use radio 
images at 8640 and 4800 MHz made with the Australia Telescope Compact 
Array (ATCA) as part of a survey of the entire LMC \citep{DM03}.  These
are combinations of several short-integration samples and not as sensitive
or complete as would be gotten from a full aperture-synthesis observation 
of the SNR alone.  The 4800 MHz data have a half power beam width (HPBW) 
of 33\arcsec, while the 8640 MHz data have a HPBW of 20\arcsec; we 
convolved the latter to match the 33\arcsec\ HPBW of the former.  A 
detailed description of the ATCA continuum survey of the LMC will be 
published by \citet{D+04}.

\section{Results}

\subsection{Morphology}

\begin{figure}[b]
\epsscale{1.0}

\plotone{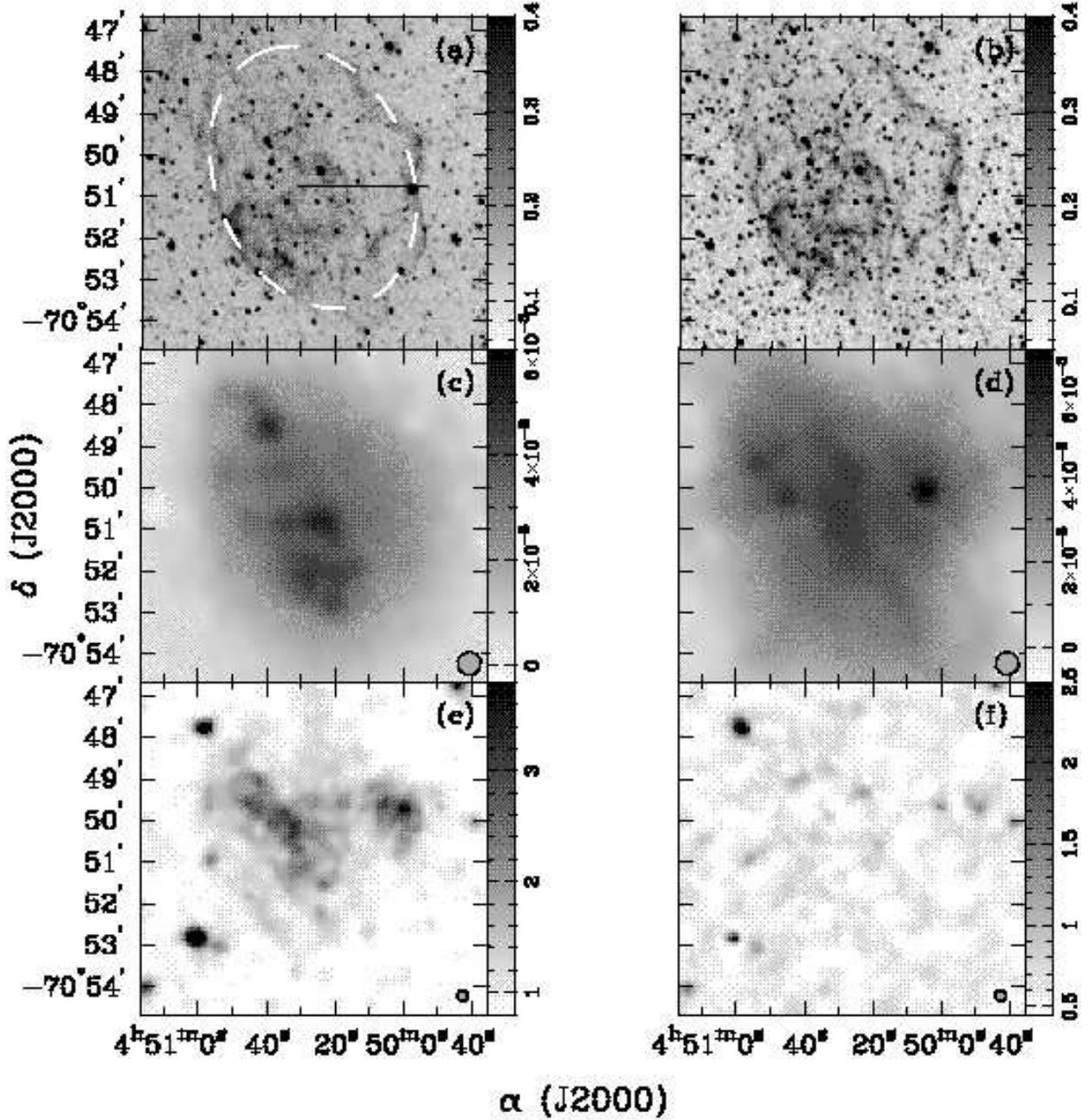}
\caption{Images of \snr\ on same spatial scales, displayed with a linear 
greyscale. (a) CTIO-MCELS \ha\ (in units of counts s$^{-1}$ pixel$^{-1}$).  
The dashed ellipse is the estimated optical extent; the solid line is the 
location of the echelle slit. (b) CTIO-MCELS \sii\ (as in a) (c) ATCA 
radio 4800 MHz (Jy/beam) (d) ATCA radio 8640 MHz (as in c) (e) \xmm\ 
MOS 0.2-2.0 keV (counts) adaptively smoothed to S/N ratio 6. (f) \xmm\ 
MOS 2.0-8.0 keV (as in e). Beam sizes for c-d, and average PSF 
for e-f, are shown in the lower right-hand corner of the panels.}
\label{fig:allband_img}
\end{figure}

The MCELS images show the \ha\ emission from \snr\ to be clearly shell-like, 
if irregular, against a diffuse background, much as reported by \citet{M+85}.  
The shell appears slightly more limb-brightened along the eastern and western 
sides.  Some filamentary emission also appears interior to this shell (at 
least in projection). In \sii, the shell appears more brightly against the 
diffuse background and the interior filaments are the brightest section of 
the SNR (Fig.~\ref{fig:allband_img}a,b).

\begin{figure}[b]
\epsscale{1.0}
\plotone{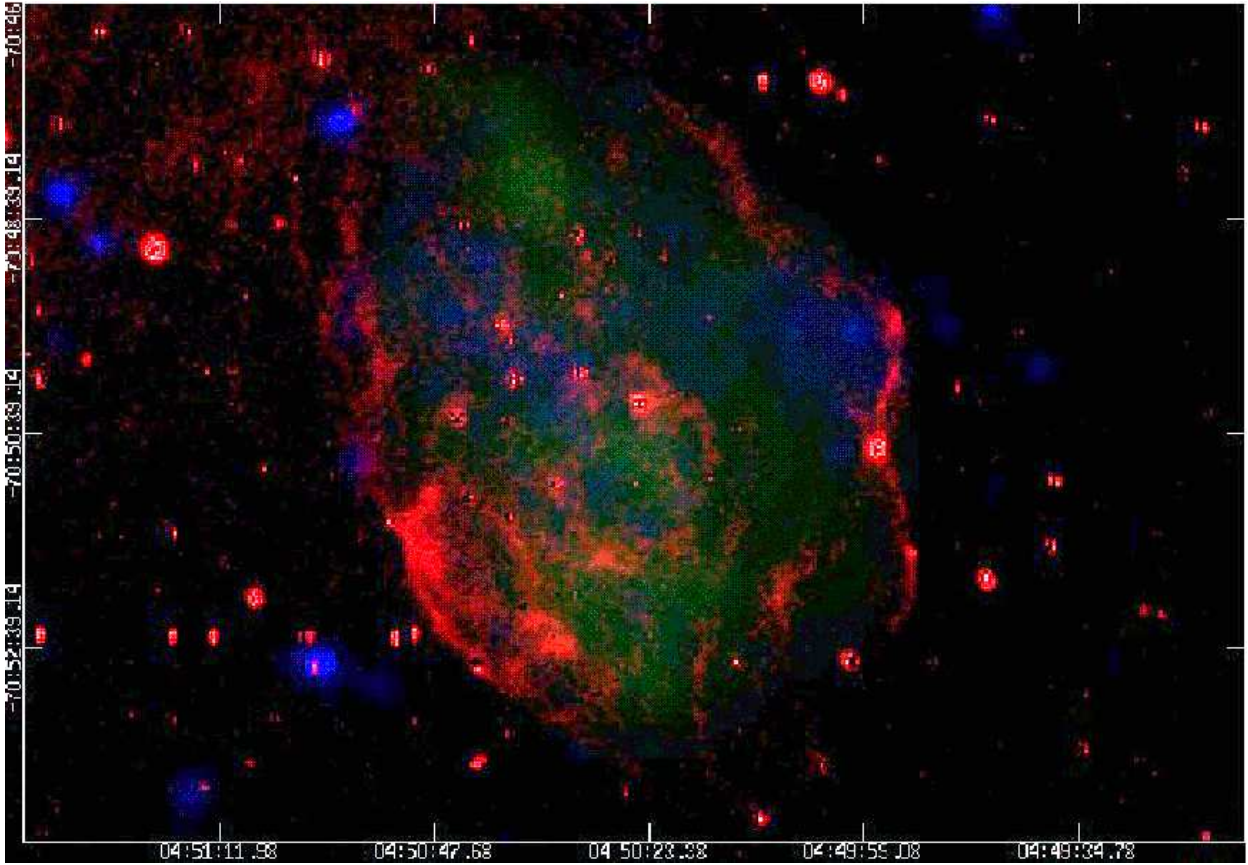}
\caption{Three-waveband image, showing spatial correspondance: 
(red) CTIO-MCELS \ha, (green) ATCA radio 6 cm,
(blue) \xmm\ MOS (merged) 0.2-3.0 keV, adaptively smoothed.}
\label{fig:3col}
\end{figure}

\snr\ was clearly detected in the individual \xmm\ EPIC MOS images.  The
emission covered a roughly elliptical area with major and minor axes of 
6\farcm0 and 4\farcm6. In the merged image (Fig.~\ref{fig:allband_img}e), 
faint emission is slightly more visible above the background level, including 
an arc along most of the eastern side of the SNR, corresponding well to the 
shell seen in optical images. Toward the northern and southwestern ends of 
the SNR, however, the X-ray emission still does not reach the full optical 
extent. Throughout the X-ray emitting region, the surface brightness 
increases toward the SNR center, with the exception of a bright spot at 
a protrusion on the western side of the SNR.  The X-ray morphology falls 
between the ``centrally brightened" and ``diffuse face" categories used 
in the classifications of \citet{W+99}.  \citet{RP98} define a category
of SNRs, which show centrally-filled X-ray emission combined with a radio
shell,  as ``mixed morphology."  Here, we expand that definition to 
include SNRs with centrally-filled thermal X-ray emission in a shell
remnant observed at any wavelength, which would describe \snr\  
(Fig.~\ref{fig:3col}).

Surprisingly, the radio images do not show shell-like emission. In the 
4800 MHz image, diffuse radio emission can be seen over the face of the 
entire remnant, with possible bright patches along the major axis of the 
SNR.  The outline of the noisy 8640 MHz image more closely resembles the 
X-ray distribution, with broad peaks corresponding to the X-ray maxima in 
the center and along the western limb.  A centrally filled radio morphology 
can indicate the presence of a pulsar, but no point source is detected in 
either radio or X-ray observations.

\subsection {Physical Structure}

\subsubsection{Hot Interior}

\begin{figure}[b]
\epsscale{1.0}
\plotone{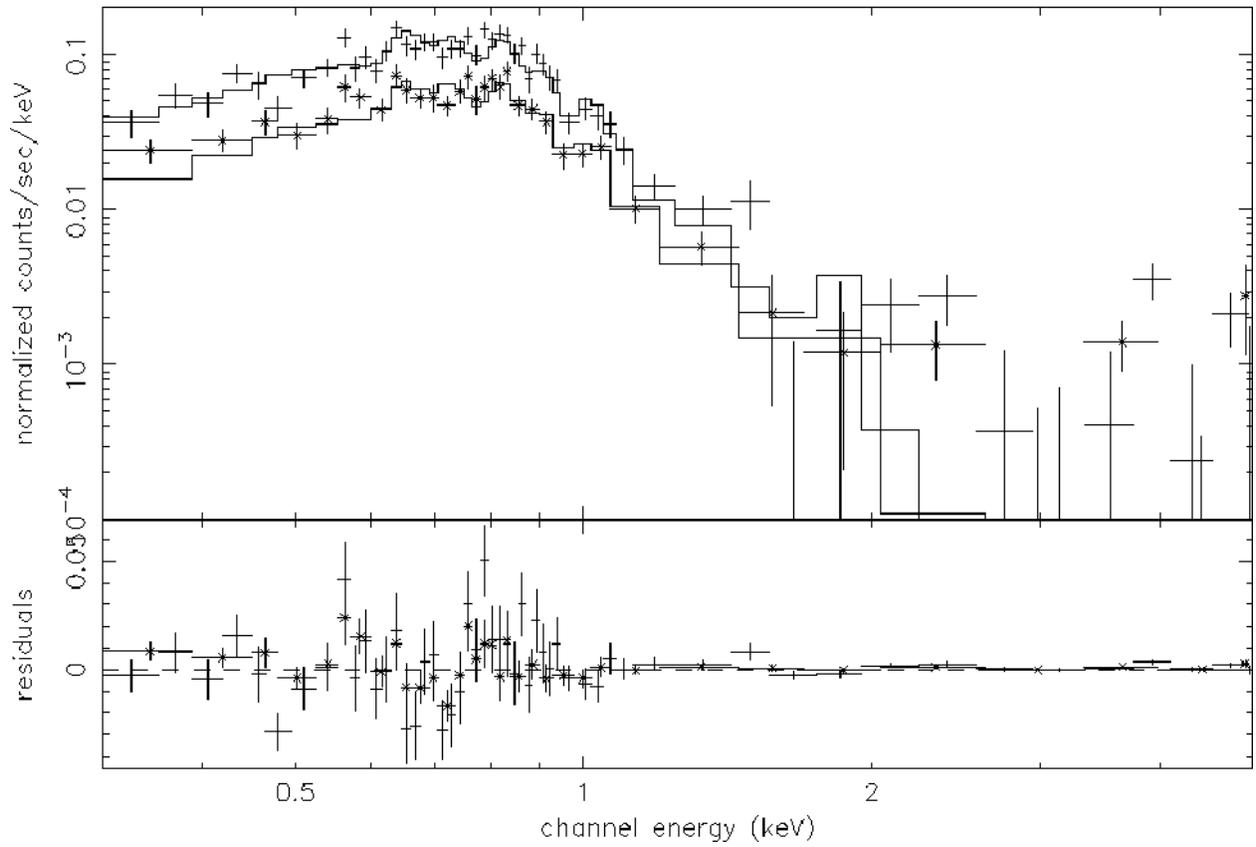}
\caption{\xmm\ MOS spectra and best spectral fits. Upper (+) points and 
line show the data and fit of both MOS detectors, merged; lower (*) 
points and line show the data and fit for the MOS1 detector only.}
\label{fig:xmmspec}
\end{figure}

The \xmm\ EPIC collected 5684 source counts with MOS1 and 5763 source 
counts with MOS2 for \snr\ over the total selected good-time intervals of
48.8 ks.  The majority of the X-ray  emission from the SNR's interior is 
below 2 keV, and is consistent with emission from a thermal plasma.  Other 
models, such as a simple power-law, can be ruled out at the 90\% confidence 
level.  Given the advanced evolutionary stage of the SNR, we expect its
age to be sufficiently close to the ionization timescale to neglect the
effects of nonequilibrium ionization; this fact and the low X-ray count
rate led us to choose a simple equilibrium-ionization plasma model. 
A MeKaL\footnote{Details and references for the MeKaL thermal plasma
model can be found at \\
\url{http://heasarc.gsfc.nasa.gov/docs/xanadu/xspec/manual/node39.html}} 
model fit gave parameters as follows: absorption column density of 
$N_{\rm H} = 1.1\pm 0.9\times10^{21}$ cm$^{-2}$, plasma temperature 
$kT=0.28\pm0.04$ keV, abundances of $0.14\pm0.1$ times solar values, 
and a normalization factor\footnote{Normalization factor 
$A ({\rm cm}^{-5}) = 10^{-14} \int n_{\rm e}^2 dV/4\pi D^2$, where 
$n_{\rm e}$ is the electron density, $V$ is the volume, and $D$ is the 
distance, all in cgs units.}
of $A$=$8\pm4 \times10^{-4}$ cm$^{-5}$. Errors given are representative 
of the range of values at the 90\% confidence level, and were determined
by varying the free parameters together using the \textsc{error} command 
in \textsc{xspec}.  Although technically the error of the normalization 
is interdependent with those of the other parameters, we have given only 
the uncorrelated component, as this parameter is varied during the fitting
process of \textsc{error}. The quoted errors are statistical, 
and do not include systematic contributions from, e.g., the possible 
presence of more than one temperature component.  While the fit is 
statistically acceptable (reduced $\chi^2=1.08$ for 337 degrees of freedom), 
the error bars for the energy bins allow considerable latitude for such 
fits.  The best-fit spectrum for the merged event file was very similar to 
that for the individual MOS spectra (Fig.~\ref{fig:xmmspec}). Based on this 
model fit, we obtain an absorbed flux of 1.7$\times10^{-13}$ 
erg~cm$^{-2}$~s$^{-1}$, or an unabsorbed flux of 3.9$\times10^{-13}$ 
erg~cm$^{-2}$~s$^{-1}$, over the 0.2 to 10 keV range. At a distance of 
50 kpc, this equates to a luminosity of 1.2$\times10^{35}$ erg~s$^{-1}$.

As a check on our fitted value of $N_{\rm H}$, we determined the amount of 
Galactic $N_{\rm H}$ toward \snr\ from the work of \citet{DL90}, which gave 
an average value of 9.0 $\times$ 10$^{20}$ cm$^{-2}$.  To this, we added 
the estimate of \citet{R+84} for the internal column density of the LMC
along the closest line of sight to \snr, 7.4 $\times$ 10$^{20}$ cm$^{-2}$.
Although the measurement of $N_{\rm H}$ for the LMC includes material behind 
\snr\ as well as in front of it, it does not include contributions from 
other molecular or ionized gas. The two sources of error tend to offset 
one another, so we treat these factors as negligible in our estimation.  
The resulting estimate of total $N_{\rm H}$ = 1.6 $\times$ 10$^{21}$ cm$^{-2}$ 
is within the 90\% error range of our fitted value.

The normalization factor of the MeKaL model can be used to determine the 
density of the hot gas. We calculate the volume of the hot gas by 
measuring the optical and X-ray extent of the SNR, treating each as an 
ellipsoid with a height equal to the average of the semimajor and semiminor 
axes that describe its extent on the sky.  We assume that the hot gas 
occupies any area not filled by the cool shell, and vice versa. We measure 
a linear difference between the optical and X-ray extent of about 2 pc, 
implying a shell thickness of approximately that width; the volume of 
hot gas comes to 8$\times 10^{60}$ cm$^3$.   Within this X-ray emitting 
volume, we assume a volume filling factor of 1, to reflect the centrally 
filled X-ray morphology of the remnant.   We further assume a distance of 
50 kpc for the SNR, and that hydrogen and helium are fully ionized 
($n_{\rm e} \sim 1.2 n_{\rm H}$).  Then, using the normalization given above, 
we find a density in the hot gas of $n_{\rm H}=0.06\pm0.03$ cm$^{-3}$.  This 
low density is unsurprising, as the smaller extent of the X-ray emission, 
compared to the optical, suggests that most of the radiation is coming from 
the cavity evacuated during the earlier SNR expansion.  Using these numbers 
for the hot gas volume and density, we find a total mass of hot gas of 
8 $\pm$ 2 $\times 10^{35}$ g, or 400 $\pm$ 100 M$_{\sun}$.  Quoted errors 
are simply propagation of the random errors in the fitted quantities.

These estimates of volume and density can be used to calculate the 
thermal energy and thermal pressure in this hot gas, from 
$E_{\rm th} = \frac{3}{2} nVkT$ and $P = nkT$, respectively. In both 
cases we use the value of the temperature derived from spectral fits, 
0.28 keV.  This yields values of $E_{\rm th}$ = 3.2 $\pm$ 0.9 $\times 
10^{50}$ erg and $P$ = 2.7 $\pm$ 0.7 $\times 10^{-11}$ dyne~cm$^{-2}$.

\subsubsection{Cool Shell}

\begin{figure}[b]
\epsscale{0.75}
\plotone{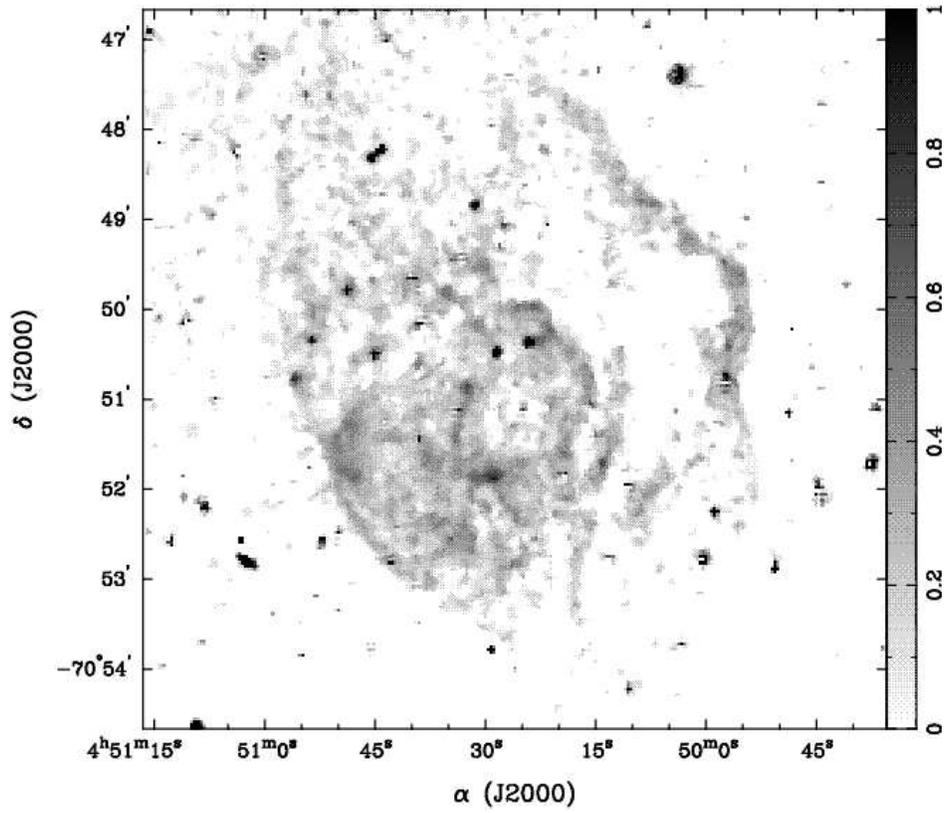}
\caption{\sii/\ha\ ratio map, based on CTIO MCELS images. Image is displayed
on a linear scale with an intensity range of ratios from 0 to 1.}
\label{fig:ratio}
\end{figure}

The shell of cool material behind the shock is mostly visible at optical 
wavelengths, with recently cooled shocked gas delineated by areas of 
higher \sii/\ha\ ratios. As shown in Figure~\ref{fig:ratio}, the \sii/\ha\ 
ratios in many optical filaments are relatively low for an SNR, only 
exceeding 0.6 in a few patches, most of which are located in a roughly 
circular region in the remnant's interior.  This inner region corresponds 
well, spatially, to the small area of clearly nonthermal emission seen in 
radio (\S 3.2.3). It also corresponds well to a region of clear red-shifted 
expansion seen in echelle spectroscopy as described below.

\begin{figure}[b]
\epsscale{1.0}
\plotone{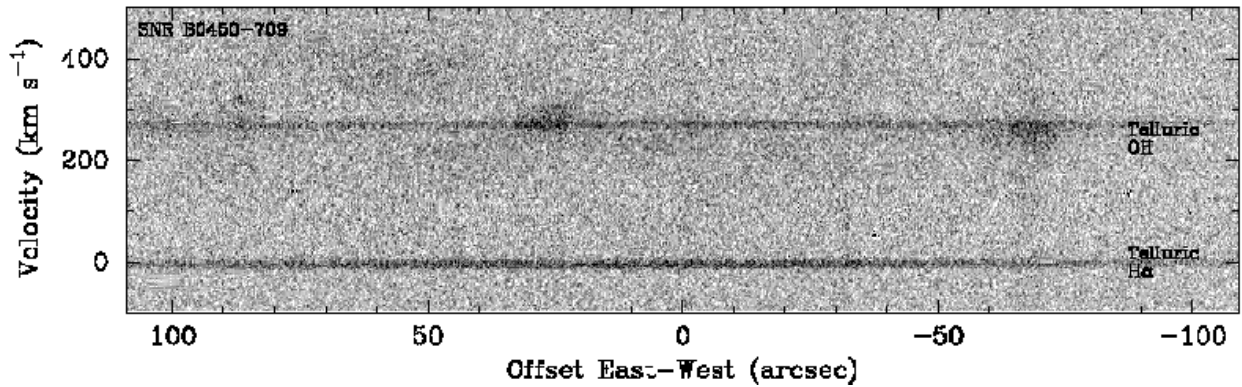}
\caption{Echelle spectrogram (Fig.~\ref{fig:allband_img} shows slit position).
The vertical axis shows heliocentric velocities. The geocoronal \ha\ line at 
6563\AA\ is marked.  The telluric OH line appears very close to the line
of Doppler-shifted \ha\ from LMC gas toward \snr.}
\label{fig:echelle}
\end{figure}

The optical echelle data show several lines in the \ha\ spectral region. 
The Doppler-shifted nebular \ha\ emission toward the SNR  
includes both a velocity component constant along the slit, representing 
the background interstellar gas and adopted as the SNR's systemic velocity,  
and multiple regions of emission deviating from this systemic velocity, 
showing motions within the expanding gas.  Also detected are 
the narrow geocoronal \ha\ (6562.85 \AA) and telluric OH 6-1 P2(3.5) 
6568.779 lines \citep{O+96}, both of which are constant along the slit 
(Fig.~\ref{fig:echelle}). 

Emission at the systemic velocity of \snr\ is faint and unfortunately 
overlaps with that from the telluric OH line, making this velocity more 
difficult to discern  with high accuracy. However, a plot of the velocity 
profile in one of the regions of brighter SNR emission, in which the data
are summed over 25\arcsec\ along the slit,  appears to show two components, 
one at 6568.6 \AA\ and the other the telluric OH line at 6568.7 \AA. We 
identify the former as the Doppler-shifted component representing the 
systemic velocity of the SNR, which would imply a heliocentric velocity 
of 271 $\pm$3 km~s$^{-1}$. Accordingly, the shifts in velocity 
of expanding material from the SNR shell ($\Delta v$) are 
measured from this estimated systemic velocity.

While there is some resemblance to the classic bow-shaped pattern of a 
rapidly expanding shell, the distribution of the velocities is uneven, 
with multiple occurrences of convergence toward, and deviation from, the 
systemic velocity.  Of particular interest is the fact that there appears
to be a section across the SNR's face which shows a distinct expansion 
pattern in red-shifted emission, with a fainter counterpart at 
blue-shifted velocities.   Spatially, this velocity pattern 
lies along the circular region of enhanced \sii/\ha\ emission and
nonthermal radio emission, centered at R.A. 04$^h$50$^m$23$^s$ and
$\sim$90\arcsec\ in diameter.  Elsewhere along the slit, the SNR shows 
maximum velocities of  $\Delta v_{\rm blue}$ = $-$120 $\pm$ 5 km~s$^{-1}$ 
and $\Delta v_{\rm red}$ = $+$52 $\pm$ 5 km~s$^{-1}$. However, in the 
region of the distinct expansion pattern, maximum velocities are  
$\Delta v_{\rm blue}$ = $-$130 $\pm$ 10 km~s$^{-1}$ and 
$\Delta v_{\rm red}$ = $+$220 $\pm$ 10 km~s$^{-1}$. These calculations 
should of course be approached with caution, as for an expanding shell, 
it is expected that the maximum  expansion velocities will be measured at
the center of the remnant's face, where the velocity component along the 
line-of-sight is greatest.  It is the abrupt transition from somewhat
chaotic low-velocity distribution to a clear expansion pattern toward
the center that makes this region notable.

The filaments measured from the flux-calibrated MCELS \ha\ images 
have an average surface brightness of 2.0 $\pm$ 0.5 $\times 10^{-16}$ 
erg~cm$^{-2}$~s$^{-1}$~arcsec$^{-2}$, which gives an emission
measure of 104 $\pm$ 23 cm$^{-6}$ pc at a presumed temperature of 
10$^4$ K. For comparison, the background \ha\ emission at the 2$\sigma$ 
level is 8.8$\times 10^{-17}$ erg~cm$^{-2}$~s$^{-1}$~arcsec$^{-2}$, or 
an emission measure of 46 cm$^{-6}$~pc. As the emission measure is 
proportional to $\int_0^{\cal L}  n_e^2 dl$, we can calculate a density 
from the longest line-of-sight through the shell. We estimate a radius 
of 42 pc and a shell thickness of 8 $\pm$ 2\arcsec\ (2 $\pm$ 0.5 pc) in 
excellent agreement with the difference between optical and X-ray extent 
found above. Using ${\cal L} = 2 \sqrt{R^2 - (R - \Delta R)^2}$ (where 
$\Delta R$ is the shell thickness), we calculate ${\cal L}$ = 25.6 pc 
and therefore an average shell density of 2.0 $\pm$ 1.0 cm$^{-3}$. This 
figure should be treated with caution, as it is based on a spherical shell 
(as is manifestly not the case here), and as the estimate of shell thickness 
is close to the resolution of the image (2\arcsec $\times$ 2\arcsec\ pixels). 
Our estimated density is an average throughout the SNR shell, and is 
broadly typical of such average densities found in other LMC SNR shells
\citep[e.g.,][]{M+96}; individual filament and clump densities may be  
considerably higher. 

As a cross-check on our figures above, we calculate the density based on 
the \ha\ luminosity over the entire SNR, rather than from the surface 
brightness of individual filaments along the limb.  This luminosity, 
$L_{H\alpha}$, is related to the emission coefficient $j_{H\alpha}$ 
according to

\[ L_{H\alpha} = \int 4\pi j_{H\alpha} dV \]

\noindent where $V$ is the emitting volume.  We assume a temperature 
of $T$=10$^4$ K, for which $J_{H\alpha}$ = 2.4 $\times$ 10$^{-25}$ 
$n_{\rm H}$ $n_{\rm e}$  erg~s$^{-1}$~cm$^3$ \citep{O89}. We also 
assume the emitting gas consists of singly ionized hydrogen and helium 
with a number ratio of H:He = 10:1.  Over the face of the remnant, 
we measure $L_{H\alpha} \approx$ 4 $\pm$ 2 $\times 10^{36}$ erg~s$^{-1}$.
Recalling our estimate for the cool gas volume as an ellipsoidal shell 
of 2 pc thickness,  we calculate a volume for this region of 
1 $\times 10^{60}$ cm$^3$, or an approximate fractional volume  of 0.1 
for the shell with respect to the entire SNR volume.  This estimate 
yields a density of 1.4 $\pm$ 1.0 cm$^{-3}$, in good agreement with the 
value found above.  Note that this value is dependent on the filling 
factor of \ha-emitting gas within the shell, and increases as the inverse 
square root of that factor.

A density of 2.0 cm$^{-3}$ as calculated above, along with our estimate 
of the shell volume, implies a mass of 3.6 $\pm$ 1.8 $\times 10^{36}$ g, 
or 1800 $\pm$ 900 M$_{\sun}$, in the shell.  If we add this mass to that of
the hot gas, and subtract 20 M$_{\sun}$ for the progenitor, we find 
a mean value for the density in the pre-SN ISM over this volume of 
0.3 cm$^{-3}$.  For comparison, if we assume our measured shell density 
represents a factor of four compression behind the shock, the ISM 
density would be 0.5 cm$^{-3}$, consistent within the errors and 
assumptions.

Using this mass and the maximum expansion velocity (220 km~s$^{-1}$) found 
from echelle observations, we can use $E_{\rm kinetic} = \onehalf M_{\rm 
shell}  v_{\rm exp}^2$ to derive a kinetic energy of 9 $\pm$ 4 $\times 
10^{50}$ erg.  If instead we assume that the 220 km~s$^{-1}$ velocity is 
associated with a separate shell, and use the maximum velocity measured in 
the outer filaments (120 km~s$^{-1}$) to represent the expansion velocity 
of the SNR, we obtain a kinetic energy of 3 $\pm$ 1 $\times 10^{50}$ erg.  
Likewise, we can calculate the thermal pressure in the cool shell 
from $P=nkT$, using our calculated density and an estimated shell
temperature of 10$^4$ K; we find a value of 2.8 $\pm$ 1.4 $\times 10^{-12}$
dyne~cm$^{-2}$.  As this is still an order of magnitude less than the 
value derived for the hot gas, we suggest that this SNR may still be
pressure-driven, in agreement with the current understanding of the 
expansion of older remnants \citep[e.g.,][and references therein]{BP04}.
This remnant is among the few old SNRs whose cavity and shell pressures 
can be determined sufficiently for comparison with model predictions. 
This is a significant finding, therefore, as it helps to confirm
observationally the analytic and numerical finding that interior 
pressure is a significant factor in the expansion of SNRs to late ages.

\subsubsection{Radio Emission}

\begin{figure}[b]
\epsscale{0.9}
\plotone{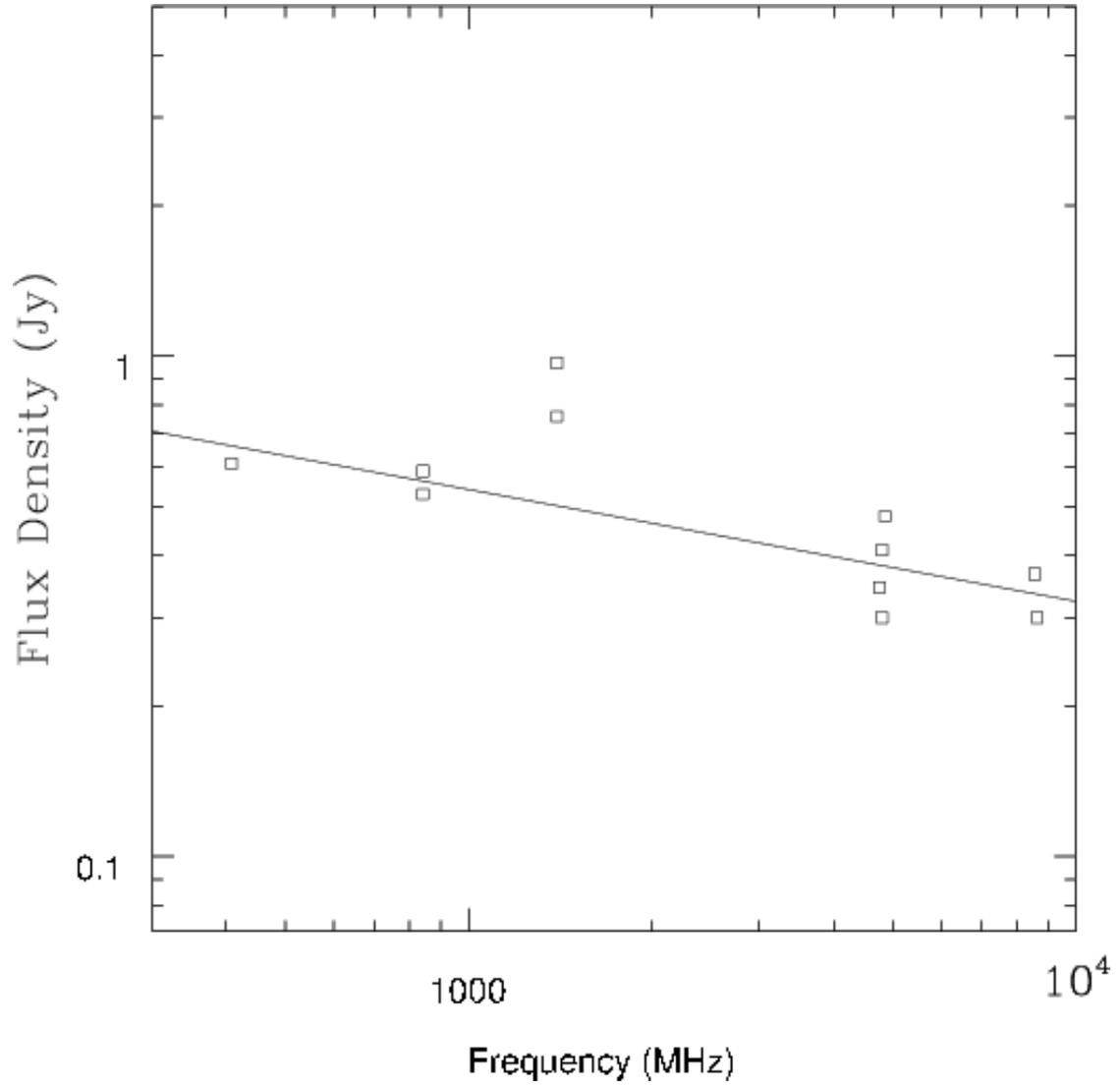}
\caption{Radio spectral index from flux density values
as given in Table~\ref{tab:radindx}.}
\label{fig:radindex}
\end{figure}

The radio spectrum of \snr\ is relatively flat; only the central 6-cm peak 
(which also corresponds to the bright \sii/\ha\ region in the optical) can 
be confidently described as nonthermal, although the best-fit spectrum for 
the entire remnant is steeper than that of thermal radiation.  We combined 
our flux density values with those of previous radio observations at various 
frequencies \citep{MBB72,CLM76,M+85,F+95,BLS99} and used a linear regression 
fit (log frequency vs. log flux density) to obtain a spectral index for the 
SNR of $-$0.21 (Table~\ref{tab:radindx}, Fig.~\ref{fig:radindex}). Note that 
1480 MHz data are included in the plot, but not used in the fit, as these 
data have insufficient resolution to exclude emission from another nearby 
bright source to the southwest. The spectral index for the SNR as a whole 
is unusually flat compared to the synchrotron spectrum expected for 
shell SNRs, typically $-$0.5\footnote{Trushkin, S.~A. 1999, 
\url{http://cats.sao.ru/snr\_spectra.html}}; but radio spectral indices 
are known to flatten with shell age \citep{CS84}. We detect several spots 
of polarized emission at the 3$\sigma$\ level only around the periphery of 
the remnant at 4.8 GHz, but they indicate approximately 30 -- 50 \% 
polarization and so are probably unreliable.

To demonstrate that the radio emission is in fact dominantly nonthermal, 
we compare the emission in \ha\ to its equivalent in the radio.  If the 
radio emission were entirely thermal, the expected 6 cm flux density could 
be calculated from the ratio of the radio emission coefficient to the \ha\ 
emission coefficient. Following the derivation in \citet{CD86}, we find

\[ \frac{j_\alpha}{j_\nu} 
 =  8.608 \times 10^{-10} \frac{1}{1 + (N_{\rm He}/N_{\rm H})} 
 \left( \frac{T}{10^4} \right) ^{-0.59}
 \left( \frac{\nu}{10^9} \right) ^{0.1} \]

\noindent where $j_\alpha/j_\nu$ is in erg cm$^{-2}$ s$^{-1}$ Jy$^{-1}$, 
$T$ is in K, and $\nu$ is in Hz. Taking $N_{\rm He}/N_{\rm H} = 0.1$, 
$T=10^4$ K, and $\nu = 4800$ MHz thus gives 
$j_\alpha/j_\nu = 9.15 \times 10^{-10}$ erg cm$^{-2}$ s$^{-1}$ Jy$^{-1}$.
This ratio is directly related to the ratio of the \ha\ and radio fluxes:

\[ \frac{F({\rm H}\alpha)}{S_{\nu} ({\rm radio})} 
 = \frac{j_\alpha}{j_\nu} 10^{-0.4 A_{\alpha}} \]

We presume that, given the small extinctions involved, we can reasonably 
approximate $A_{\alpha}$ = $A_{\rm V}$.  The visual extinction $A_{\rm V}$ 
can be calculated from the absorption column density obtained from the X-ray 
fits, using the LMC value of 2 $\times$ 10$^{22}$ atoms cm$^{-2}$ mag$^{-1}$
from \citet{K82} to obtain $A_{\rm V} \approx 0.16$ mag.  Using the column
density value calculated from \hi\ observations \citep{DL90,R+84} gives a 
slightly higher $A_{\rm V} \approx 0.25$ mag.  In either case, the 
contribution of reddening in the visible band is negligible.  Using the \ha\ 
flux given above, and correcting for the difference between the 6 cm beam 
size and the \ha\ pixel size, we find an expected 
$S_{\nu} = 0.31$ mJy~beam$^{-1}$ if thermal emission is the sole source 
of the observed 6 cm flux.  Given that the actual measured surface brightness 
ranges from 1 to 5 mJy beam$^{-1}$ over the face of the remnant, we conclude 
that a substantial fraction of the radio emission is nonthermal.

The emission at radio wavelengths within SNRs is usually generated at the 
outer edge of the expanding shock (and, when applicable, the reverse shock), 
where accelerated relativistic electrons encounter compressed magnetic 
fields and thus generate synchrotron emission. Hence, one usually expects 
a shell-like distribution of nonthermal radio emission, in contrast to 
the distributed, relatively flat-spectral-index emission we observe in 
\snr.  In older remnants, the radio morphology tends to match the optical 
because the emission is generated in the compression of cooler filaments 
\citep{DV75}.  However, this object may be old enough that some of the 
emission has diffused toward the center and much of the bright radio 
emission corresponds to the central object seen in bright \sii/\ha\ 
filaments.

\section{Discussion}

Given the aspherical morphology and large extent of this SNR, it is 
difficult to calculate its probable age.  If one takes an average radius 
of 42 pc, and estimates an expansion velocity of 220 km~s$^{-1}$ using 
the maximum expansion seen from the echelle data, one can use the relation 
$t=\eta R / V_{exp}$ for a rough estimate.  Assuming point-blast expansion
\citep[$\eta$=0.4]{S59} gives an estimated age of 75,000 yr. However, it 
is quite possible that a remnant of this size, unless the external ISM is 
very tenuous, is well into the radiative, momentum-conserving 
phase. \citet{BP04} have developed an analytic solution for 
a momentum-conserving SNR for which interior thermal pressure is still 
a significant factor. In their picture, $\eta$ (there called $m$) 
will gradually fall from the value derived in numerical calculations 
($\eta \sim 0.33$) to the asymptotic value of 
$\eta = \frac{2}{7} = 0.286$.  Using the former value and the
parameters above yields an age of 62,000 yr, while the latter 
value gives 53,000 yr.  The optical and X-ray properties described 
above are consistent with a SNR in this age range.

\citet{SC92} and \citet{S99} predict that, as a remnant ages, the expansion 
velocity slows to the point where X-rays are no longer generated by 
the shock front.  The remaining X-ray emission is from the hot interior 
of the SNR, where material had been shocked to high temperatures by the 
previous phases of more rapid SNR expansion.  Thus, the dominant source
for the remaining X-ray emission is from ``fossil radiation" due to this
hot interior, and the SNR appears centrally filled in X-rays.
This has been extended as a partial explanation for ``mixed morphology"
SNRs, although other physical processes, such as thermal conduction or
evaporation of dense cloudlets, must be invoked to explain the high 
surface brightness of this extremely tenuous gas \citep{RP98}.
This ``hot bubble" is expected to shrink with respect to the full 
extent of the SNR as it ages, due to the cooling of the gas near
the surface of the ``hot bubble". The extension of X-ray emission in 
\snr\ to the western edge of the optical shell, and the echelle-derived 
expansion velocity, suggests that in some regions, the shock front is 
still capable of producing soft X-rays.  However, the centrally filled 
morphology and the overall smaller extent of the X-ray emitting region 
than the radio/optical SNR may indicate that not only has fossil 
radiation begun to predominate, but cooling of the ``hot bubble" in some 
areas may be well underway.  Alternately, the regions of low X-ray 
emission may simply be too faint for detection above noise levels, 
even with this long \xmm\ exposure.  

To test whether this picture of \snr\ was consistent with our observations, 
we compared our findings with spectra simulated from nonequilibrium 
ionization (NEI) hydrocode for a SNR in a relatively low-density region 
with significant ambient nonthermal pressure.  A detailed description of 
the SNR model is given by \citet{S99}.  The particular simulation used 
here assumed an explosion energy of 10$^{51}$ erg, ambient density of 
0.1 cm$^{-3}$, and an effective ambient magnetic field  of 5 $\mu$G.  We 
followed the SNR development up to an age of 3 $\times$ 10$^6$ yr, showing 
the evolution of the soft (0.1-2.2 keV) X-ray spectrum over this period.  
We applied a multiplicative model for photoelectric absorption to this 
range of model spectra, and fit them to our \xmm\ data. 

The model fits did not provide a unique solution, but rather showed
several local minima in the $\chi^2$ fits within a fairly restricted
region of parameter space. The values of reduced $\chi^2$ (1.4 for 
110 degrees of freedom) for these minima are statistically significant 
only at very low confidence levels, but are at least reasonable given 
the amount of noise in the data. These ``best fits" were obtained for 
ages of 45,000 yr through 95,000 yr, with column densities of 2-4  
$\times$ 10$^{21}$ cm$^{-2}$, somewhat higher than the absorption found 
for a simple plasma model.  This range of ages, though broad, is 
consistent with the range of kinematically-derived ages,  providing 
an independent consistency check to our age derivations above.

To estimate the effects of the ambient density on the model results, we
also compared our data to modeled spectra for a remnant in an ambient 
ISM of 0.01 cm$^{-3}$ \citep[``SNR C" from][]{S99}.  Unsurprisingly, the
best fits occurred at the high end of the age range (95,000 yr), and 
also required higher absorption column densities of $\sim$5 $\times$ 
10$^{21}$ cm$^{-2}$.  This is consistent with our expectation that 
a remnant within a lower ambient density medium would evolve and cool 
more slowly than its counterpart in a denser ISM.   The fact that the 
kinematically derived ages are more consistent with the models for a 
0.1 cm$^{-3}$ ISM than for a 0.01 cm$^{-3}$ ISM provides some additional
support of the range of ambient densities (0.3-0.5 cm$^{-3}$) calculated
above for the medium around \snr.

These estimates for the advanced age of this SNR also provide a 
framework for the interpretation of the radio morphology. At this age, 
the shock strength may be low enough that relativistic particles are no 
longer accelerated at the outer edges of the SNR and so the radio 
synchrotron emission is distributed more evenly over the remnant.
We can then attribute the lack of a clear shell to the 
fact that the SNR has passed into a late-evolutionary stage, and 
therefore is subject to volume emissivity enhancements as in IC 443
\citep{DV75}, following the distribution of cool neutral material.
This gas is expected to occupy (at a small volume filling factor) 
a wide region behind the shock, which would lead to a more distributed
radio surface brightness. In addition, if \snr\ is expanding into a
clumpy ISM as suggested by the irregular morphology, the distribution
of such material - and thus the enhancement to radio emission - may be
further spread out across the remnant, leading to a ``patchy" radio
surface brightness as seen in \snr.

The lack of a steeper spectral index for the radio emission, one of the 
classic signatures of SNRs, is puzzling. The relativistic particles 
responsible for most of the  radio synchrotron emission have lifetimes
ranging from 10$^5$ to 10$^7$ yr, so even if there were no new 
acceleration of such particles, we would still expect the existing 
supply of relativistic electrons to exhibit significant nonthermal radio 
emission with a spectral index closer to $-$0.5.   For example, \snr\ 
shares certain characteristics with the Galactic SNR W28. \citet{RB02} 
found that, as with \snr, the X-ray emission from W28 is unevenly 
distributed, elliptical on the sky, and centrally brightened.  Much of 
W28 appears to be in the radiative stage, with expansion velocities 
$<$100 km~s$^{-1}$. However, W28 still has a spectral index of $-$0.4 
\citep{K92}, steeper than that seen in \snr, although it also possesses 
a ``flat spectrum core".  The latter, however, is commonly thought to 
result from a pulsar-wind nebula (PWN) contribution, which is unlikely 
in the case of \snr.  It would be quite unusual for a remnant of \snr's
age to contain a PWN, and the extent of the emission (diameter $>$70 pc) 
would make it over twice as large as the largest currently known Crab-type 
SNR \citep{GDG00}.  Furthermore, no regions of hard ($>$ 3 keV) X-ray 
emission are seen within \snr, which argues against the presence of
any region of nonthermal X-ray emission as would be expected from a PWN.
It is possible that the lack of a well-determined nonthermal radio emission 
spectrum is simply a function of the faintness of the SNR emission with 
respect to the background, making it difficult to accurately determine 
the spectral index.  

Our study of the X-ray emission from \snr, in conjunction with radio and
optical data, provides a relatively consistent overall scenario for this 
SNR, but one which leaves open several possibilities concerning the 
development of the interior substructure.  Our estimates of the physical
parameters of the SNR (hot gas, shell, and ambient densities; thermal
pressures; thermal and kinetic energies) are consistent with the picture
of a large, old shell in a late stage of SNR evolution, with its development
largely a factor of shell momentum and interior pressure.  However, the
distributed radio emission is not a commonly seen phenomenon, and therefore 
leaves open the question of whether this too is a natural consequence of 
late-stage SNR expansion.  

The detection of significant X-ray emission over the face of this SNR 
has also provided us with a rare opportunity to directly test models 
of late-stage SNR development against the data.  Although our preliminary
analysis allows wide latitude for error, we are certainly able to show 
that one such set of models is consistent with the data for reasonable
choices of physical parameters. Future work drawing on the sensitivity
of instruments such as those on \xmm\ will enable the discovery, analysis,
and rigorous comparison with models for the highly underrepresented 
population of well-evolved remnants such as \snr.

\acknowledgements
The authors thank the anonymous referee for very helpful comments. 
RMW, YHC, and JRD acknowledge support from NASA grant NAG 5-11159.

\begin{deluxetable}{cccccc}
\tablecaption{Radio Flux Observations}
\tablehead{
\colhead{Freq} &
\colhead{Flux} &
\colhead{Err} &
\colhead{Telescope} &
\colhead{Citation} \\
\colhead{(MHz)} &
\colhead{(Jy)} &
\colhead{(Jy)} &
\colhead{} &
\colhead{} 
}
\startdata
4800  &  0.41   &  0.15 & Parkes &  \citet{MBB72} \\
 408  &  0.61   &  0.04 & Molonglo Cross &  \citet{CLM76} \\
 843  &  0.530  & \null & MOST &  \citet{M+85} \\
4750  &  0.345  &  0.04 & Parkes &  \citet{F+95} \\
4850  &  0.479  &  0.05 & Parkes &  \citet{F+95} \\
8550  &  0.367  &  0.06 & Parkes &  \citet{F+95} \\
8640  &  0.3    &  0.1  & ATCA &  this work \\
4800  &  0.3    &  0.1  & ATCA &  this work \\
 843  &  0.590  & \null & MOST &  \citet{BLS99} \\
1400  &  0.97   & \null & Parkes &  \citet{MBB72} \\
1400  &  0.758  &  0.13 & Parkes &  \citet{F+95} \\
\enddata
\label{tab:radindx}
\end{deluxetable}

\begin{thebibliography}{}

\bibitem[Bandiera \& Petruk(2004)]{BP04}Bandiera, R., \& Petruk, O. 2004, 
	A\&A, in press, astro-ph/0402598
\bibitem[Bock, Large, \& Sadler(1999)]{BLS99}Bock, D.~C.-J., Large, M.~I.,
	\& Sadler, E.~M. 1999, AJ, 117, 1578
\bibitem[Caplan \& Deharveng(1986)]{CD86}Caplan, J. \& Deharveng, L. 1986, 
	A\&A, 155, 297  
\bibitem[Chu et al.(2003)]{Chuetal03} Chu, Y.-H., Chen, C.-H.~R.,
	Danforth, C., Dunne, B.~C., Gruendl, R.~A., Naz{\' e}, Y., Oey,
	M.~S., \& Points, S.~D.\ 2003, \aj, 125, 2098
\bibitem[Chu \& Kennicutt(1988)]{CK88}Chu, Y.-H., \& Kennicutt, R.~C.~Jr.\  
	1988, AJ, 96, 1874
\bibitem[Chu et al.(2000)]{C+00}Chu, Y.-H., Kim, S., Points, S.~D., Petre, R. 
	\& Snowden, S. L. 2000, AJ, 119, 2242
\bibitem[Chu \& Mac Low(1990)]{CM90}Chu, Y.-H., \& Mac Low, M.~M. 1990, 
	ApJ, 365, 510
\bibitem[Clarke, Little, \& Mills(1976)]{CLM76} Clarke, J.~N., Little, A.~G., 
	\& Mills, B.~Y. 1976,  AuJPA, 40, 1
\bibitem[Cowsik \& Sarkar(1984)]{CS84}Cowsik, R. \& Sarkar, S. 1984, 
	MNRAS, 207, 745
\bibitem[Dickel \& McIntyre(2003)]{DM03}Dickel, J.~R., \& McIntyre, V. 2003,
	BAAS, 202, 4001
\bibitem[Dickel et al.(2004)]{D+04}Dickel, J.~R., McIntyre, V., Gruendl, R.~A.,  
	\& Milne, D.~K. 2004, AJ, to be submitted
\bibitem[Dickey \& Lockman(1990)]{DL90}Dickey, J.~M., \& Lockman, F.~J.
	1990, ARAA, 28, 215	
\bibitem[Dodorico \& Sabbadin(1976)]{DS76}Dodorico, S. \& Sabbadin, F. 1976,
	A\&A, 53, 44
\bibitem[Duin \&  van der Laan(1974)]{DV75}Duin, R.~M., \&  van der Laan, H. 
	1975, A\&A, 40, 111
\bibitem[Feast(1999)]{F99} Feast, M. 1999, IAUS, 190, 542
\bibitem[Filipovic et al.(1995)]{F+95}Filipovic, M. D., Haynes, R.~F., 
	White, G.~L., Jones, P.~A., Klein, U., \& Wielebinski, R. 1995,
	A\&AS, 111, 311
\bibitem[Gaensler, Dickel, \& Green(2000)]{GDG00}Gaensler, B.~M.,
	Dickel, J.~R., \& Green, A.~J. 2000, ApJ, 542, 380
\bibitem[Kassim(1992)]{K92}Kassim, N.~E. 1992, AJ, 103, 943
\bibitem[Koornneef(1982)]{K82}Koornneef, J. 1982, A\&A, 107, 247
\bibitem[Lazendic, Dickel, \& Jones(2003)]{LDJ03}Lazendic, J.~S., 
	Dickel, J.~R., \& Jones, P.~A. 2003, ApJ, 596, 287
\bibitem[Lowry et al.(2004)]{Lowryetal04} Lowry, J.~D., Chu, Y.-H.,
	Guerrero, M.~A., Gruendl, R.~A., Snowden, S.~L., \& Smith, R.~C.\
	2004, \aj, 127, 125
\bibitem[Mathewson et al.(1985)]{M+85}Mathewson, D. S., Ford, V. L., Tuohy, 
	I. R., Mills, B. Y., Turtle, A. J., \& Helfand, D. J. 1985, 
	ApJS, 58, 197
\bibitem[McGee, Brooks, \& Batchelor(1972)]{MBB72}McGee, R.~X., Brooks, J.~W., 
	\& Batchelor, R. A. 1972, AuJPh, 25, 581
\bibitem[Morse et al.(1996)]{M+96}Morse, J.~A., Blair, W.~P., Dopita, M.~A., 
	Hughes, J.~P., Kirshner, R.~P., Long, K.~S., Raymond, J.~C., 
	Sutherland, R.~S., \& Winkler, P.~F. 1996, AJ, 112, 509
\bibitem[Osterbrock et al.(1996)]{O+96}Osterbrock, D.~E., Fulbright, J.~P., 
	Martel, A.~R., Keane, M.~J., Trager, S.~C., \& Basri, G. 1996, 
	PASP, 108, 277
\bibitem[Osterbrock(1989)]{O89}Osterbrock, D.~E. 1989, {\it Astrophysics of
	Gaseous Nebulae and AGN} (California: University Science Books)
\bibitem[Rho \& Petre(1998)]{RP98}Rho, J., \& Petre, R. 1998, ApJL, 503, 167
\bibitem[Rho \& Borkowski(2002)]{RB02}Rho, J., \& Borkowski, K.~J. 2002, 
	ApJ, 575, 201
\bibitem[Rohlfs et al.(1984)]{R+84}Rohlfs, K., Kreitschmann, J., 
	Feitzinger, J.~V., \&  Siegman, B.~C. 1984, A\&A, 137, 343
\bibitem[Sedov(1959)]{S59}Sedov, L.~I. 1959, {\it Similarity and Dimensional 
	Methods in Mechanics} (New York: Academic)
\bibitem[Shelton(1999)]{S99}Shelton, R. 1999, ApJ, 521, 217
\bibitem[Slavin \& Cox(1992)]{SC92} Slavin, J.~D., \& Cox,
	D.~P.\ 1992, \apj, 392, 131
\bibitem[Slavin \& Cox(1993)]{SC93} Slavin, J.~D., \& Cox,
	D.~P.\ 1993, \apj, 417, 187
\bibitem[Smith et al.(1999)]{Smithetal99}Smith, R.~C., \& The 
	MCELS Team 1999, IAU Symp.~190: New Views of the Magellanic Clouds, 
	190, 28
\bibitem[Williams et al.(1999)]{W+99} Williams, R.~M., Chu, Y.-H. Dickel, 
	J.~R., Petre, R., Smith, R.~C., \& Tavarez, M. 1999, ApJS, 123, 467
\end{thebibliography}
\end{document}